\documentclass[a4paper ,12pt]{article}

\usepackage{graphicx}
\usepackage{amsfonts}
\usepackage{amsthm}
\usepackage{amssymb,amsmath}
\usepackage[usenames,dvipsnames]{color}
\usepackage{hyperref}
\hypersetup{
    pdftitle={Courant-like brackets and loop spaces},    
    pdfauthor={ Joel Ekstrand and Maxim Zabzine},     
    pdfnewwindow=true,      
    colorlinks=true,       
    linkcolor=black,          
    citecolor= black,        
    filecolor= black,      
    urlcolor= black           
}
\usepackage{microtype} 
\numberwithin{equation}{section}

\newcommand{\Section}[1]{\section{#1} \setcounter{equation}{0}}
\newcommand{\beq}{\begin{equation}}
\newcommand{\eeq}[1]{\label{#1}\end{equation}}
\newcommand{\ber}{\begin{eqnarray}}
\newcommand{\eer}[1]{\label{#1}\end{eqnarray}}

\newcommand{\ud}{\mathrm{d}}
\newcommand{\uD}{\mathrm{D}}
\newcommand{\ui}{\mathrm{\iota}}

\def\+{{+\!\!\!+}}

\newcommand{\eg}{\textit{e.g.}~}
\newcommand{\ie}{\textit{i.e.\ }}

\newcommand \intsloop{\hspace{-0.5em}  \int\limits_{T[1]S^1} \hspace{-0.5em}  \ud\sigma \ud \theta~}

\begin{document}
\renewcommand{\theequation}{\thesection.\arabic{equation}}
\setcounter{page}{0}
\thispagestyle{empty}
\begin{flushright} \small
UUITP-09/09 \\ 
\end{flushright}
\bigskip

\bigskip
\begin{center}
\Huge
{\bf{ Courant-like brackets\\ and loop spaces}}
 \\[12mm]
  \normalsize
{\bf Joel Ekstrand and Maxim Zabzine} \\
\bigskip
{\small\it
 Department of Physics and Astronomy, 
Uppsala University, \\ Box 516, SE-751\,20 Uppsala, Sweden}
\end{center}
\bigskip
\begin{abstract}
\noindent
We study the algebra of local functionals equipped with a Poisson bracket. We discuss the underlying  algebraic structures related to a version of the Courant-Dorfman algebra. 
As a main illustration,  we consider the functionals over the  cotangent bundle of the superloop space
  over a smooth manifold.  We present a number of examples of the Courant-like brackets arising from this analysis. 
 \end{abstract}
 
 \eject
\normalsize



\eject

\section{Introduction}
\label{introduction}

Higher algebraic  structures have been a subject of recent interest in physics and geometry.    
 In this work, we study  the algebraic structures underlying the Poisson brackets of local functionals in classical  field theory.  

 Our work is inspired by two different observations. 
  The first interesting observation was made in 1997, in \cite{Barnich:1997ij}, 
   about the appearance  of an  $L_\infty$-structure on the space  of local functionals in classical field theory (see also \cite{ AlAshhab:2003th, markl}).
     The work \cite{Barnich:1997ij} is a formalization of the simple idea that a local functional is an integral over some function (integrand),  and all operations with the local functionals involve throwing away total  derivatives.  Thus, the Poisson brackets  of local functionals induce a binary operation on the integrands which satisfy the properties of Poisson brackets up to total derivatives. We thus end up with the appropriate $L_\infty$-structure on the space of local functionals in classical field theory.  Later, another curious observation was made in  \cite{Alekseev:2004np}. Namely, the authors found
          an intriguing  way  to 
  ''derive'' the Courant and Dorfman brackets on $TM\oplus T^*M$ by calculating the classical Poisson brackets  between local functionals of a special form defined over the cotangent bundle of a loop space\footnote{ Similar observations were made in \cite{Bonelli:2005ti} (see  \cite{FigueroaO'Farrill:2005uz}
             for relevant  comments)  and more elaborated in \cite{Guttenberg:2006zi}.}. 
The Courant bracket by itself  is an example of an $L_\infty$-structure \cite{roytenbergweinst}.  It would be natural that these 
 two observations are related and that they are just different manifestations of the same algebraic structures. 
 
 In the present work, 
  we argue that the appropriate formalization of the Courant-Dorfman structure naturally appears when one 
   considers the algebra of local functionals. 
  For the sake of concreteness, we consider the case of the string phase space (\ie the cotangent bundle of the loop space) 
 and its supersymmetric  generalization. However, many results can easily be generalized beyond this concrete framework. 
  Beside arguing about the general structure, we are able to generate infinitely many examples of Leibniz algebras for 
   appropriate geometric objects over a smooth manifold $M$. We present four concrete examples. 
 
 The paper is organized as follows. In section \ref{Alekseev-Strobl} we review the observation made by Alekseev and Strobl from 
  \cite{Alekseev:2004np}. We also discuss its supersymmetric generalization and we set the conventions for the rest 
   of the paper.  Section \ref{Linfinity} presents the main observation: the algebra of local functionals 
     naturally is equipped with a Leibniz bracket and there is a structure of a weak Courant-Dorfman algebra. 
    In section \ref{newcurrents} we go through different examples of this structure. 
      Section \ref{summary} contains a summary of the paper and the concluding remarks. 
    For the reader's convenience,  in the appendix  we review and summarize the key properties of the standard Dorfman and 
   Courant brackets on $TM\oplus T^*M$ and we also give the definitions of  (weak) Courant-Dorfman algebras. 

\section{The Alekseev-Strobl observation}
\label{Alekseev-Strobl}

In this section, we introduce the relevant notation and review the observation by Alekseev and Strobl 
 from \cite{Alekseev:2004np}.  Moreover, we present its supersymmetric generalization. 
 
  The bosonic string phase space can be constructed as follows. Let us consider the loop space
  $$ LM = \{ ~X : S^1 \longrightarrow M ~ \}~,$$
   which is the space of 
    differentiable maps from the circle $S^1$ to a smooth manifold $M$. 
     The cotangent bundle of  $LM$ can be defined as the space of differentiable vector bundle morphisms 
     $(X, p): TS^1 \rightarrow T^*M$ with  differentiable base maps $X : S^1 \rightarrow M$.
 The symplectic structure on $T^*LM$ is of the standard form and it 
  can be written in local coordinates  as follows:
 \beq
  \omega = \int\limits_{S^1} \ud\sigma~ \delta X^\mu \wedge \delta p_\mu~,
 \eeq{bosonicsymplectic}
  where $\delta$ is the de Rham differential on $T^*LM$.  Thus, we have a Poisson algebra
   on the space of functionals, $C^\infty (T^*LM)$.  For a section $(v+ \omega)$ of $TM\oplus T^*M$,
    we define  a local functional of the special form
    \beq
    J_{\epsilon} (v +\omega ) = \int\limits_{S^1} \ud\sigma~ \epsilon (v^\mu p_\mu + \omega_\mu \partial X^\mu )~,
    \eeq{bosoniccurrent}
  where $\partial$ is a derivative along a loop and $\epsilon : S^1 \rightarrow \mathbb{R}$ is a test function
   ($\epsilon \in C^\infty(S^1)$).
   It has been observed in \cite{Alekseev:2004np} that the Poisson bracket of these special local 
    functionals can be written as
  \beq
  \{ J_{\epsilon_1} (A) , J_{\epsilon_2} (B) \} = - J_{\epsilon_1 \epsilon_2} ( A * B ) -
   \int\limits_{S^1} \ud\sigma~ (\epsilon_2 \partial \epsilon_1) \langle A, B
   \rangle ~,
 \eeq{ASexpressionDorfman}
  or, alternatively, as
 \beq
  \{ J_{\epsilon_1} (A) , J_{\epsilon_2} (B) \} = - J_{\epsilon_1 \epsilon_2} ([A, B]_C) +
   \int\limits_{S^1} \ud\sigma~ (\epsilon_1 \partial \epsilon_2 -
   \epsilon_2 \partial \epsilon_1) \langle A, B  \rangle ~.
 \eeq{ASexpressionCourant}
 In these expressions $A,B \in \Gamma(TM\oplus T^*M)$, $*$ stands for the Dorfman bracket, 
  $[~,~]_C$ for the Courant bracket and $\langle~, ~\rangle$ is the natural pairing on $TM\oplus T^*M$
   (see the appendix for a review). 
 
 This setup can easily be generalized to the supersymmetric case.  Let us define the superloop 
  space as the space of the following maps:
  $${\cal L} M = \{~ \Phi : T[1]S^1 \longrightarrow M~\}~,$$
   where $T[1]S^1$ is a superloop parametrized by the coordinates $(\sigma, \theta)$, with $\sigma$ 
    being a coordinate along $S^1$ as before and $\theta$ is its partner which is a section of the cotangent 
     bundle to the circle with reversed parity.  In local coordinates  $\Phi$ is 
         \beq  
     \Phi^\mu (\sigma , \theta ) = X^\mu(\sigma) + \theta~ \lambda^\mu(\sigma)~.
      \eeq{expansionofPhi}
  Therefore,  the superloop space can  alternatively be defined as
  $${\cal L}M = \{~ S^1 \longrightarrow T[1]M~\}~.$$
  The corresponding phase space is $T^*{\cal L}M$ which can be defined as the space of bundle morphisms 
  $(\Phi, S) : T[1]S^1 \rightarrow T^*[1]M$.  $T^*{\cal L}M$ is equipped with a canonical symplectic structure 
  \beq
   \omega = i  \intsloop \delta S_\mu \wedge \delta \Phi^\mu~, 
  \eeq{susysymplecticform}
   where $S$ is a coordinate along the fiber of the bundle:  
 \beq
  S_\mu (\sigma, \theta) = \rho_\mu (\sigma) + i \theta p_\mu (\sigma)~.
   \eeq{expansionofS}  
     It is important to note that the fiber is odd. The ''super''  momenta $S$ therefore anticommute: $S_\mu   S_\nu = - S_\nu   S_\mu$.
    Upon integration over $\theta$ the bosonic part of (\ref{susysymplecticform}) coincides with (\ref{bosonicsymplectic}).  
    The symplectic structure (\ref{susysymplecticform}) makes  $C^\infty (T^*{\cal L}M)$ into a super-Poisson algebra. 
     Defining the left and right functional derivatives of a functional $F(\Phi, S)$ as follows:
     \beq
     \begin{split}
          \delta F &= \intsloop  \left ( \frac{ F \overleftarrow{\delta}}{\delta S_\mu} \delta S^\mu +
  \frac{F \overleftarrow{\delta}}{\delta \Phi^\mu} \delta \Phi^\mu \right ) \\ 
  &= \intsloop \left ( \delta S_\mu \frac{ \overrightarrow{\delta} F}{\delta S_\mu}  +
  \delta \Phi^\mu \frac{\overrightarrow{\delta} F}{\delta \Phi^\mu}  \right ) ~,
     \end{split}
     \eeq{leftrightderivative}
     we end up  with the corresponding super-Poisson bracket:
\beq
\{ F, G \} = i \intsloop \left ( \frac{ F \overleftarrow{\delta}}{\delta S_\mu} \frac{\overrightarrow{\delta} G}{\delta \Phi^\mu} - \frac{F \overleftarrow{\delta}}{\delta \Phi^\mu} \frac{\overrightarrow{\delta} G}{\delta S_\mu} \right )~.
\eeq{susybracketgeneral} 
In what follows, we will often drop the prefix ''super'', the meaning will hopefully still be clear.
  As in the bosonic case, we can define a local functional associated to a section of $TM\oplus  T^*M$ as follows:
  \beq
   J_\epsilon (v + \omega) = \intsloop  \epsilon \left ( v^\mu S_\mu +  \omega_\mu \uD \Phi^\mu \right )~,
  \eeq{susycurrentaaa}
  with $\epsilon (\sigma, \theta)$ being an even test function. The derivative $\uD$ is defined as 
\begin{equation} \label{oddDerivDef}
\uD= \frac{\partial}{\partial \theta} + i \theta \partial ~. 
\end{equation}
So, we have one even derivative, $\partial$,  and one odd derivative, $\uD$, with
\begin{equation} \label{DDequalsiPartial}
\uD^2 = \uD  \uD = i \partial~.
\end{equation}
The factor $i$ in the definition \eqref{oddDerivDef} is a conventional choice, in order to make $\uD^2$ an hermitian operator. Also note that $\uD \Phi^\mu \uD \Phi^\nu = - \uD \Phi^\nu \uD \Phi^\mu$. 

 With respect to the bracket (\ref{susybracketgeneral}) the local functionals (\ref{susycurrentaaa}) satisfy 
  \beq
   \{ J_{\epsilon_1} (A) , J_{\epsilon_2} (B) \} = i J_{\epsilon_1 \epsilon_2} ([A, B]_C)
   +\frac{i}{2} \intsloop  (
   \epsilon_2  \uD \epsilon_1  - \epsilon_1 \uD \epsilon_2) \langle A, B  \rangle ~.
 \eeq{susyAlekseevStrobl}
 This  is the supersymmetric generalization of the bosonic bracket \eqref{ASexpressionCourant}, which has been
   discussed  previously in \cite{Guttenberg:2006zi}. 

The space $T^*{\cal L}M$ can be equipped with a more general symplectic structure than 
 \eqref{susysymplecticform}. These symplectic structures are labelled by a closed three form on $M$.  All results and observations 
  can easily be generalized to this situation. However, for clarity, we avoid the case of the most 
   general symplectic structure on $T^*{\cal L}M$.  Further details about $T^*{\cal L}M$ can be found 
    in \cite{Bredthauer:2006hf, Zabzine:2006uz}.

\section{Leibniz algebra on local operators}
\label{Linfinity}
For the sake of clarity, we first discuss the bosonic case. The supersymmetric generalization would be straightforward, and 
  we will comment on it later on. The space of smooth functionals $C^\infty (T^*LM)$ is a Poisson algebra.  In physics, however,  
   the local functionals play a special role. A local functional is defined as follows:
\beq
 J_\epsilon (A) = \int\limits_{S^1}\ud\sigma ~\epsilon(\sigma) A (X, \partial X, \ldots, \partial^k X, p, \partial p, \ldots, \partial^l p)~,
\eeq{localfunctionalbosonic}
 where $\epsilon$ is a test function and $A$ is a function of fields and their derivatives (only a finite number of derivatives are allowed).  
 The local functionals  form a subalgebra with respect to the Poisson bracket. This subalgebra is not a Poisson subalgebra though, since we  cannot define a product of two local functionals as a local functional.  Next, we define a binary operation $*$ 
  on the local operators as follows:
\beq
 \{ J_1(A), J_\epsilon (B) \} = J_\epsilon (A * B)~.
\eeq{basicDorfman}
 Combining the Jacobi identity
\begin{equation}\label{Jacobi}\begin{split}
& \{ J_1(A), \{ J_1 (B), J_\epsilon (C)\}\} + \\ 
 & \{ J_\epsilon (C), \{ J_1 (A), J_1 (B)\}\} + \\ 
&\{ J_1(B), \{ J_\epsilon (C), J_1 (A)\}\} =0 
\end{split}\end{equation}
   and the antisymmetry of the Poisson bracket 
  with the definition \eqref{basicDorfman}, we arrive at the relation 
  \beq
   J_{\epsilon} \left( \; A * ( B* C) - B* (A * C) - (A*B) *C \; \right) = 0~.
  \eeq{currentsatisfiedaa}
   Since (\ref{currentsatisfiedaa}) should be true for any test function $\epsilon$, we get
 \beq
A * ( B* C) = (A*B) *C + B* (A * C) ~.
\eeq{Leinbnitz}
 The binary operation $*$  thus gives rise to a Leibniz algebra. A (left) Leibniz algebra (sometimes called a Loday algebra) is a module  over a commutative ring or field (in our case $\mathbb{R}$)
  with a bilinear product $*$ such that \eqref{Leinbnitz} is satisfied. 
    In other words, left multiplication by any element $A$ is a derivation.
 Using the definition \eqref{basicDorfman}, we write the Poisson bracket between two local functionals, with arbitrary test functions, in 
  the following form:
 \begin{subequations} \label{fullanswerassanomalyterm}
\begin{equation} \label{fullanswerass}
  \{ J_{\epsilon_1} (A), J_{\epsilon_2} (B) \} = J_{\epsilon_1\epsilon_2} (A * B) +  \Lambda (\epsilon_1, \epsilon_2 ; ~A, B)~,
\end{equation}
where the last term $\Lambda$ will be referred to as \textit{the anomalous term}.  The general form of the anomalous term is 
 \beq
  \Lambda (\epsilon_1, \epsilon_2 ; ~A, B)= \sum\limits_{i=1}^{\infty}~ \int\limits_{S^1} \ud\sigma~ (\epsilon_2 \partial^{(i)}\epsilon_1)
  ~ f_i (A, B)~,
 \eeq{anomalyterm} 
 \end{subequations}
  where $f_i(A, B)$ are expressions  constructed out of $A$ and $B$. 
 For concrete $A$ and $B$, the sum in (\ref{anomalyterm}) would only have a finite number of terms.  We conclude that the Poisson  bracket between any two local functionals can be represented in the form \eqref{fullanswerassanomalyterm}, thus inducing 
   a Leibniz algebra structure on the local operators.  It is important to stress that the decomposition on the right hand of \eqref{fullanswerass}
     into a $J$-term and an anomalous
    term a priori is not unique. This ambiguity is removed  by requiring that $\Lambda (1, \epsilon_2; ~A, B)=0$, in agreement with our definition \eqref{basicDorfman}.  
      The Alekseev-Strobl formula \eqref{ASexpressionDorfman} is just a particular  example of this general 
     calculation \eqref{fullanswerassanomalyterm}.

In the context of quantum field theory, a more familiar form of  \eqref{fullanswerassanomalyterm}  is
\beq
     \{ A(\sigma), B(\sigma')\} = (A*B)(\sigma') \delta(\sigma-\sigma') + \sum\limits_{i=1}^{\infty} f_i (A, B)(\sigma') ~\partial^{(i)}_{\sigma'} \delta (\sigma' -\sigma)~.
    \eeq{deltafunctionform}
This is to be understood as an equality of distributions, yielding \eqref{fullanswerassanomalyterm} upon multiplying with test functions and integrating.

The antisymmetry of the Poisson bracket implies
  \beq
   J_{\epsilon_1 \epsilon_2} (A * B + B*A) = - \Lambda (\epsilon_1, \epsilon_2; ~A, B) -  \Lambda (\epsilon_2, \epsilon_1; ~B, A)~.
  \eeq{antisymmetry}
    Using the prescription $\Lambda (1, \epsilon_2; ~A, B)=0$ together with the property that (\ref{antisymmetry}) is true 
     for any choice of the test functions we arrive at 
  \beq
   A* B + B* A = \sum\limits_{i=1}^{\infty} (-1)^{i-1} \partial^{(i)} f_i (A, B)= \sum\limits_{i=1}^{\infty} (-1)^{i-1} \partial^{(i)} f_i (B, A) ~. 
  \eeq{naturalpairinga} 
   Let us define a symmetric bilinear form by
  \beq
 \langle A, B\rangle \equiv \frac{1}{2}\sum\limits_{i=1}^{\infty} (-1)^{i-1} \partial^{(i-1)}  \left ( f_i (A, B)+  f_i (B, A) \right )~,
\eeq{najsj29292}
which allows us to rewrite equation \eqref{naturalpairinga} as
 \beq
   A* B + B* A = \partial  \langle A, B\rangle~.
 \eeq{dhsjje999}
     Moreover, using the definition (\ref{basicDorfman}) and the property $J_\epsilon (\partial A) = - J_{\partial \epsilon} (A)$, 
      we obtain the following relation between $\partial$ and $*$:
      \beq
      \partial A * B =0~.
        \eeq{pAstarB}
 Thus on the space of local functionals we get three operations $*, \partial, \langle ~, ~\rangle$
  which satisfy the properties (\ref{Leinbnitz}),  (\ref{dhsjje999}) and (\ref{pAstarB}). 
   In the above discussion of local functionals, we have ignored the properties of the $f_i$-operations. 
       Imposing the Jacobi identity on the expressions \eqref{fullanswerassanomalyterm}
       would give, in general,  infinity many relations between the operations $*$ and $f_i$. 
       In each concrete calculation, only a finite number of these relations will be non trivial, due to the assumption of the functionals being local.
This is very reminiscent of the Poisson vertex algebra structure. Thus, in all generality, we should have been discussing a sheaf of Poisson vertex algebras associated to M.
 The structures we are investigating here  is only the tip of the iceberg if we adopt this general point of view.     
  However, this restricted structures still lead to highly non-trivial geometrical results, as can be seen from our 
   examples.

Now, let us discuss possible interpretations of these structures and their relation to other works, in 
 particular to  \cite{Barnich:1997ij}.  
 It is useful to compare our manipulations with the notion of local functionals within the variational bi-complex. 
   On the space of expressions $A (X, \partial X, \ldots, \partial^k X, p, \partial p, \ldots, \partial^l p)$
      we  have two natural operations: variation and taking the full derivative along the loop. These two operations 
       can be defined as two anticommuting differentials with the underlying bi-grading ${\cal A}^{(p,q)}$. 
         For further details, the reader may consult  \cite{dickeybook}.  Here we closely follow the setting 
          in  \cite{Barnich:1997ij}.
          Let us define ${\cal A}^{(0,0)}$ to be 
          the zero-forms and ${\cal A}^{(1,0)}$ to be one-forms on the loop.  There is a differential~$\ud_h$:
 \beq
 {\mathbb R} ~ \longrightarrow~ {\cal A}^{(0,0)}~ \stackrel{\ud_h}{\longrightarrow} ~{\cal A}^{(1,0)}~,
\eeq{sequenceamm}
where $\ud_h f = (\partial f ) \ud\sigma$, with $\partial$ being a full derivative along the loop. The space of local functionals in the variational bi-complex is 
 ${\cal A}_{loc} = {\cal A}^{(1,0)}/\ud_h  ({\cal A}^{(0,0)})$. In 
 our definition \eqref{localfunctionalbosonic} of a local functional, which includes a test function, we use an element from ${\cal A}^{(1,0)}$.
  Therefore, the previous discussion can be formalized in the following way. 
 The space ${\cal A}^{(1,0)}$ is equipped with a bracket~$* : {\cal A}^{(1,0)} \times {\cal A}^{(1,0)} \rightarrow {\cal A}^{(1,0)}$ 
  and an inner product $\langle~,~\rangle : {\cal A}^{(1,0)} \times {\cal A}^{(1,0)} \rightarrow {\cal A}^{(0,0)}$ such that the following 
   properties are satisfied:
    \begin{subequations}
      \label{subequations}
\begin{align}
A * ( B* C) &= B* (A * C) + (A*B) *C~,\\
  A * B + B* A  &= \ud_h \langle A, B \rangle~,\\
  (\ud_h f) * A &= 0~,
\end{align}
\end{subequations}
where $A, B, C \in {\cal A}^{(1,0)}$ and $f, g \in {\cal A}^{(0,0)}$. 
  Thus, the space of expressions ${\cal A}^{(1,0)}$ together with ${\cal A}^{(0,0)}$ form a weak Courant-Dorfman 
   algebra as defined in the appendix.  ${\cal A}_{loc}={\cal A}^{(1,0)}/\ud_h  ({\cal A}^{(0,0)})$ is equipped with a Lie bracket
    which is the Poisson bracket on the local functionals. 
   Moreover, there is a simple  $L_\infty$-structure  which has been discussed in \cite{Barnich:1997ij} (see also \cite{markl}  and 
   see \cite{dickeypaper} for specific examples) and this is related to the weak Courant-Dorfman algebra
    discussed above. 
        
The previous  discussion has a straightforward generalization to the supersymmetric case $T^*{\cal L}M$. The formulas  \eqref{fullanswerassanomalyterm} remains valid upon using the odd derivative $\uD$ instead of $\partial$, and also integrating over the odd coordinate $\theta$ in addition to the even coordinate $\sigma$.  The supersymmetric case drastically extends the space of local operators, allowing,  e.g., anti-symmetric tensors to be contracted with odd vectors. With this grading, we could consider both odd and even local functionals, leading to a $\mathbb{Z}_2$-graded Leibniz 
 algebra and a $\mathbb{Z}_2$-graded weak Dorfman-Courant algebra.  In the present work, however, we only consider even functionals.

Most of the formulas can be extended to the case of higher dimensional field theories. In particular,  the argument 
  around (\ref{basicDorfman})-(\ref{Leinbnitz}) will remain true and 
   the structure of a Leibniz algebra would be a generic feature of classical field theories.  The structure of the anomalous term 
    (\ref{anomalyterm}) will be much more involved due to the presence of derivatives in different directions. Relations like 
    (\ref{dhsjje999}) and (\ref{pAstarB}) still should be possible to generalize, but the exact form of the higher dimensional analogue to a weak Courant-Dorfman still remains to be explored.


    One can restrict the algebra of local functionals, and study functionals of a special form, which form a closed Leibniz subalgebra.
          Our arguments are also applicable to the case when we only look 
       on this Leibniz subalgebra with some specific anomalous terms. 
        In the next section we present a few interesting examples of Leibniz subalgebras arising from 
         this sort of calculations.  These examples naturally give rise to weak Courant-Dorfman algebras associated 
          to the smooth manifold $M$.

\section{New brackets from local functionals}
\label{newcurrents}

This section provides an illustration for the general considerations presented in the previous section. We consider 
 local functionals on $T^*{\cal L}M$ of special forms, which are parametrized by geometrical data on $M$. 
   These local functionals  give rise to interesting Leibniz subalgebras of 
   ${\cal A}_\text{loc}$,
    which we discuss
    here in geometrical terms.
  We recover a well-known generalization of the Courant bracket as well as new generalizations. 

In all examples except \ref{se:bracksymtensor}, it is crucial that we consider the supersymmetric phase space $T^*{\cal L}M$, since the local functionals are constructed out of anti-symmetric tensors.

\subsection{Schouten bracket}

Let us start with a simple example. Consider a local functional on $T^*{\cal L}M$ of the following form:
\begin{equation}
 J_\epsilon (v) = \intsloop \epsilon~v^{\mu_1 \ldots \mu_p} (\Phi) S_{\mu_1} \ldots S_{\mu_p} ~,
\end{equation}
where $v$ is an antisymmetric $p$-multivector field, $v \in \Gamma( \Lambda^p TM)$, and $\epsilon $ is a test function with parity $|\epsilon| = (-1)^{p+1}$. The parity of the test function is chosen in order for $ J_\epsilon (v)$ to be an even functional.
The Poisson bracket (\ref{susybracketgeneral}) between two such local functionals is given by 
\begin{equation} 
\begin{split}
\{ J_{\epsilon_1} (v)  ,  J_{\epsilon_2} (u)  \}  = & \intsloop \epsilon_1 \epsilon_2 \;  \Bigl( p~ v^{\mu_1 \ldots \mu_{p-1} \rho} \partial_\rho u^{\mu_p \ldots \mu_{p+q-1}} -\\   
&    \quad (-1)^{(p+1)(q+1)} q ~u^{\mu_1 \ldots \mu_{q-1} \rho} \partial_\rho v^{\mu_q \ldots \mu_{p+q-1}}     \Bigr) S_{\mu_1} \ldots S_{\mu_{p+q-1}}   \\
&= i J_{\epsilon_1 \epsilon_2} ([v,u]_s)~,
\end{split}
\end{equation}
where $u \in \Gamma( \Lambda^q TM)$ and $[v, u]_s$ is the Schouten bracket. 
This calculation can be extended to the local functionals parametrized by any sections of $\Lambda^\bullet TM$. 

\subsection{Courant bracket on $TM \oplus \wedge^\bullet T^*M$}

Consider the following functional:
\beq
J_\epsilon (v + \beta) = \intsloop \epsilon ~ \left  (v^\mu S_\mu + e \; \frac{1}{p!} \beta_{\nu_1 \nu_2 \ldots \nu_{p}} \uD \left  (
\uD  \Phi^{\nu_1} \uD \Phi^{\nu_2}\ldots \uD \Phi^{\nu_{p}} \right )  \right  )~,
 \eeq{eq:currentvbeta}
 where $\epsilon$ is an even test function,
   $v$ is a vector field and $\beta$ is a $p$-form. In order for the functional to have an even grading, we include $e$, a constant ( $\uD e = 0$) with parity $|e|=(-1)^{p}$. This is merely a calculational shortcut, to avoid dealing with a $\mathbb{Z}_2$-graded Poisson bracket.
The Poisson bracket between two such local functionals is  
\begin{equation} \label{eq:poissonbrJp-1Jp-1} 
\{J_{\epsilon_1} (A)  ,  J_{\epsilon_2} (B)  \}  =  i  J_{\epsilon_1 \epsilon_2}( A  \ast B ) +   \Lambda(\epsilon_1, \epsilon_2; A, B)~,
\end{equation}
where $A  \ast B$ is the Dorfman bracket  (\ref{definitionDorfman}) 
 generalized to forms of arbitrary degree.  The $\Lambda$-term is given by the following expression 
\begin{multline}
\Lambda(\epsilon_1, \epsilon_2; A, B) =  i  \intsloop \left (    e \;  \epsilon_2 \uD^2 \epsilon_1 ~
  \frac{1}{(p-1)!} \langle A, B \rangle_{\nu_1 \ldots \nu_{p-1}} \uD \Phi^{\nu_1} \ldots  \uD \Phi^{\nu_{p-1}}   \right .  \\ 
\left . -     e \; \epsilon_2 \uD \epsilon_1  ~  \frac{1}{(p-1)!}\langle A, B \rangle_{\nu_1 \ldots \nu_{p-1}}  \uD \left ( \uD \Phi^{\nu_1} \ldots  \uD \Phi^{\nu_{p-1}} \right )  \right )~ ,
\end{multline}
 where $\langle~,~\rangle$ stands for the pairing (\ref{jdjee0303}) extended to $\Gamma(TM \oplus \wedge^p T^*M)$. 

Furthermore,  if we take the test functions $\epsilon_i$ to be purely bosonic, they obey $\int \ud\theta \epsilon_i = 0$ and as a result there is
  the relation $ \int \ud\sigma\ud \theta \,\,\ \uD \epsilon_1 \uD \epsilon_2 (\text{anything}) = 0$. Using this, we can rewrite the $\Lambda$-term as
   follows:
\begin{equation}
\Lambda(\epsilon_1, \epsilon_2; A, B) = i \intsloop e \; \epsilon_2 \uD \epsilon_1  ~ \frac{1}{p!}
 \left( \ud \langle A, B \rangle \right)_{\nu_1 \ldots \nu_{p}} \uD \Phi^{\nu_1} \ldots  \uD \Phi^{\nu_{p}}  ~ .
\end{equation}
Anti-symmetrizing the expression  \eqref{eq:poissonbrJp-1Jp-1} and assuming bosonic test functions we arrive 
 at the expression
\begin{multline} 
\{ J_{\epsilon_1} (A)  , J_{\epsilon_2} (B)  \}  =  i  J_{\epsilon_1 \epsilon_2}( [A,B]_C ) \\ 
+   \frac{i }{2}  \intsloop e \; (\epsilon_2 \uD \epsilon_1  -  \epsilon_1 \uD \epsilon_2)  ~\frac{1}{p!} \left( \ud \langle A, B \rangle \right)_{\nu_1 \ldots \nu_{p}} \uD \Phi^{\nu_1} \ldots  \uD \Phi^{\nu_{p}} ~,
\end{multline}
where $[ , ]_C$ is the Courant bracket generalized to forms of any degree.

 Geometrically, we can interpret this results as follows. We can choose 
 ${\cal E} = \Gamma (TM \oplus \wedge^p T^*M)$ equipped with the Dorfman bracket $*$, and ${\cal R} = \Gamma(\wedge^{p-1}T^*M)$.
  The symmetric bilinear form $\langle~,~\rangle : {\cal E} \otimes {\cal E} \rightarrow {\cal R}$ is defined by the formula \eqref{jdjee0303} 
  and the exterior derivative is understood as $\ud : {\cal R} \rightarrow {\cal E}$.
    It can easily be checked that this structure $({\cal E}, {\cal R}, \ud, \langle~,~\rangle, *)$  satisfies the definition of 
   a weak Courant-Dorfman algebra (see the appendix for details).  
   
   With this structure, we can naturally introduce the notion of a Dirac structure ${\cal D}$ as a subbundle of $TM \oplus \wedge^p T^*M$ 
    such that $\Gamma({\cal D})$ is closed under $*$ and $\ud\langle A, B \rangle =0$ for any $A,B  \in \Gamma({\cal D})$.  
     For $A\in \Gamma({\cal D})$ the local functionals (\ref{eq:currentvbeta}) form a closed algebra under the Poisson bracket
      since $\Gamma({\cal D})$ is a Lie algebra with respect to $*$. 

\subsection{Courant-like bracket on $TM\oplus \wedge^\bullet T^*M \oplus \wedge^\bullet T^*M$}

Next, we consider a generalization of the functionals (\ref{susycurrentaaa}) and  (\ref{eq:currentvbeta}).  For a section $A=v+\beta+\gamma \in \Gamma(TM \oplus \wedge^p T^*M
 \oplus \wedge^{p+1} T^*M)$ we associate a local functional of the following form:
 \begin{equation} \label{eq:currentpform}
 \begin{split}
 J_\epsilon (A) = \intsloop\epsilon \; \left  (v^\mu S_\mu + e\frac{1}{p!}\beta_{\nu_1 \ldots \nu_{p}} \uD \left ( \uD \Phi^{\nu_1}\ldots \uD \Phi^{\nu_{p}} \right ) \right . \\ 
\left . + e \; \frac{(-1)^p}{(p+1)!} \gamma_{\nu_{1} \ldots \nu_{p+1}} \uD \Phi^{\nu_1} \ldots  \uD \Phi^{\nu_{p+1}} \right ) ,
 \end{split}
 \end{equation}
 where $e$ again is a constant with parity $|e|=(-1)^{p}$ on order to make $ J_\epsilon (A)$ even.  The Poisson bracket between two such local functionals has the form 
\begin{equation} \label{eq:poissonbrJpp-1Jpp-1}
\{ J_{\epsilon_1} (A_1 )  ,  J_{\epsilon_2} (A_2 )  \}  =  i  J_{\epsilon_1 \epsilon_2}( A_1 * A_2) +\Lambda (\epsilon_1, \epsilon_2; A_1, A_2) ~,
\end{equation}
where 
\begin{multline}
\label{eq:dorfmanlike}
(v_1 + \beta_1 + \gamma_1) * (v_2 + \beta_2 + \gamma_2)  \equiv   \{v_1, v_2\}  + \\  {\cal L}_{v_1} (\beta_2 + \gamma_2) -
\ui_{v_2} \ud (\beta_1 + \gamma_1) + 
(-1)^p  \ui_{v_2} \gamma_1
\end{multline}
is a new Dorfman bracket which differs from the standard one by the last term. One can easily check that it satisfies the Leibniz identity. 
 The $\Lambda$-term in \eqref{eq:poissonbrJpp-1Jpp-1}  is
\begin{multline}
\Lambda =  i  \intsloop \Bigl[   e \;  \epsilon_2 \uD^2 \epsilon_1  ~ \frac{1}{(p-1)!}  ( \ui_{v_1} \beta_2 + \ui_{v_2} \beta_1 )_{\nu_1 \ldots \nu_{p-1}} \uD \Phi^{\nu_1} \ldots  \uD \Phi^{\nu_{p-1}}  \\
 -e \;  \epsilon_2 \uD \epsilon_1  ~ \frac{1}{(p-1)!} ( \ui_{v_1} \beta_2 + \ui_{v_2} \beta_1 )_{\nu_1 \ldots \nu_{p-1}} \uD \left ( \uD \Phi^{\nu_1} \ldots  \uD \Phi^{\nu_{p-1}} \right ) \\
 + e \; \epsilon_2 \uD \epsilon_1  ~ \frac{(-1)^p}{p!}( \ui_{v_1} \gamma_2 + \ui_{v_2} \gamma_1 )_{\nu_1 \ldots \nu_{p}} \uD \Phi^{\nu_1} \ldots  \uD \Phi^{\nu_{p}}  \Bigr].
\end{multline}
 If we choose  the test functions to be purely bosonic, then the $\Lambda$-term can be simplified to
\begin{multline}
\Lambda =  i  \intsloop e \;  \epsilon_2 \uD \epsilon_1 
\frac{1}{p!} \Bigl(  (-1)^p ( \ui_{v_1} \gamma_2 + \ui_{v_2} \gamma_1 ) \\
+   \ud ( \ui_{v_1} \beta_2 + \ui_{v_2} \beta_1 ) \Bigr)_{\nu_1 \ldots \nu_{p}} \uD \Phi^{\nu_1} \ldots  \uD \Phi^{\nu_{p}} ~.
\label{eq:lambdabosonicbetagamma}
\end{multline}
The anti-symmetrized version of expression \ \eqref{eq:poissonbrJpp-1Jpp-1}  with  bosonic test functions is 
\begin{multline} \label{eq:poissonbrJpp-1Jpp-1AS}
\{ J_{\epsilon_1} (A_1 )  ,  J_{\epsilon_2} (A_2 )  \}  =  i  J_{\epsilon_1 \epsilon_2}( [A_1,A_2]_C) 
 + \frac{i}{2} \intsloop e \; \left(\epsilon_2 \uD \epsilon_1  -  \epsilon_1 \uD  \epsilon_2\right) \times \\
\times \frac{1}{p!}\bigl(   (-1)^p ( \ui_{v_1} \gamma_2 + \ui_{v_2} \gamma_1 ) +   \ud ( \ui_{v_1} \beta_2 + \ui_{v_2} \beta_1 ) \bigr )_{\nu_1 \ldots \nu_{p}} \uD \Phi^{\nu_1} \ldots  \uD \Phi^{\nu_{p}}~,
\end{multline}
where $[~,~]_C$ is the anti-symmetrization of the new Dorfman bracket (\ref{eq:dorfmanlike}).

 Now, let us discuss the geometrical meaning of these structures. 
 ${\cal E} = \Gamma(TM \oplus \wedge^{p} T^*M \oplus \wedge^{p+1}T^*M)$ is equipped with the Dorfman bracket $*$ given
   by (\ref{eq:dorfmanlike}).
 Define ${\cal R}=\Gamma( \wedge^{p-1} T^*M \oplus \wedge^{p}T^*M)$. The symmetric bilinear form ${\cal E}\otimes {\cal E} \rightarrow {\cal R}$
 is given by the expression
 \beq
  \langle v_1 + \beta_1 + \gamma_1, v_2 + \beta_2 + \gamma_2 \rangle = \ui_{v_1} (\beta_2 + \gamma_2) + \ui_{v_2} (\beta_1 + \gamma_1)~. 
 \eeq{djdjjdjjjj}
 Define a linear map $\ud_h : {\cal R} \rightarrow {\cal E}$ by 
\beq
\ud_h (b_{p-1} + a_p) =  \ud b_{p-1} +\ud a_p + (-1)^p a_p~,
\eeq{djdjw9999}
 where $b_{p-1}$ is $(p-1)$-form and $a_p$ is $p$-form. If we consider the operation $\ud_h$ defined for   all values values of $p$   
  by this formula, then $\ud_h^2=0$. 
The symmetrization of the bracket (\ref{eq:dorfmanlike})  is given by
\begin{equation}  
  A_{1} * A_{2} + A_{2} * A_{1} =   (-1)^p ( \ui_{v_1} \gamma_2 +  \ui_{v_2} \gamma_1) +  
   \ud \bigl( \ui_{v_1}  (\beta_2 + \gamma_2) + \ui_{v_2}  (\beta_1 + \gamma_1) \bigr) ~,
\end{equation}
 which can be written as
 \beq
 A_{1} * A_{2} + A_{2} * A_{1} = \ud_h \langle A_1, A_2 \rangle
 \eeq{eur94999}
 if we use the above definitions.  Indeed, it is easy to check that $({\cal E}, {\cal R}, \ud_h, \langle~,~\rangle, *)$ is a weak Courant-Dorfman 
  algebra.  It is interesting to mention that the bracket (\ref{eq:dorfmanlike}) can be understood as a derived bracket for the
    differential $\ud_h$ (see \cite{Derived}
   for a review of derived brackets). 
 
  The Dirac structure ${\cal D}$ would be defined as a subbundle of $TM \oplus \wedge^{p} T^*M \oplus \wedge^{p+1}T^*M$
   such that $\Gamma({\cal D})$ is closed under $*$ and $\ud_h \langle A_1, A_2 \rangle=0$ for every $A_1, A_2 \in \Gamma({\cal D})$. 
    Thus $\Gamma({\cal D})$ is a Lie algebra with respect to $*$. 
Decomposing $\ud_h \langle A_1, A_2 \rangle=0$  in form degrees give us two conditions:
\begin{align}
 \left( \ui_{v_1} \gamma_2 + \ui_{v_2} \gamma_1 \right)&= (- 1)^{p+1} \ud \bigl( \ui_{v_1}  \beta_2 + \ui_{v_2}  \beta_1 \bigr) ~,
  \label{eq:conddorfequalcour1}\\
\ud \bigl( \ui_{v_1}  \gamma_2 + \ui_{v_2}  \gamma_1 \bigr) &=0~, \label{eq:conddorfequalcour2}
\end{align}
where \eqref{eq:conddorfequalcour1} implies \eqref{eq:conddorfequalcour2}. 
Therefore it is enough that \eqref{eq:conddorfequalcour1} is satisfied. This is exactly the same as requiring that the $\Lambda$-term in \eqref{eq:lambdabosonicbetagamma} vanishes.  Thus, the local functionals (\ref{eq:currentpform}) for $A\in \Gamma({\cal D})$ 
 form a Lie algebra with respect to the Poisson bracket.

\subsection{Bracket associated with a symmetric tensor} \label{se:bracksymtensor}

Next, we consider a different example of a local functional, which is parametrized by
  a vector field~$v$, a symmetric tensor of 
 second rank~$\gamma$ and a one form~$\rho$:
\begin{equation}\label{eieieieooo}
 J_\epsilon (A) = \intsloop\epsilon \;  (v^\mu S_\mu + e\gamma_{\mu \nu} \partial \Phi^\mu \partial \Phi^\nu  + e\rho_{\mu}  \nabla \partial \Phi^\mu)~, 
\end{equation}
 where $A = v + \gamma + \rho$ and
 \beq
 \nabla \partial \Phi^\mu = \partial^2 \Phi^\mu +  \Gamma^\mu_{~\nu\rho} \partial \Phi^\nu \partial \Phi^\rho~,
 \eeq{29393000}
 with $\Gamma^\mu_{~\nu\rho}$ being a torsionless connection.   We need to introduce a connection in order to make
  the local functional \eqref{eieieieooo} invariant under diffeomorphisms of $M$.
  In \eqref{eieieieooo},  $e$ is an odd constant, and $J_\epsilon (A)$ is thus even.  
  
  The Poisson bracket between two such local functionals yields
\begin{equation}\label{ddjdkkkk}
\{ J_{\epsilon_1} (A_1 )  ,  J_{\epsilon_2} (A_2 )  \}  = i J_{\epsilon_1 \epsilon_2} (A_1 \ast A_2 )
 + \Lambda(\epsilon_1,\epsilon_1, A_1, A_2) ~.
\end{equation}
The bracket $A_1 \ast A_2$  is given by
\begin{subequations} \label{eq:symtensorbracket}
\begin{align}
(A_1 \ast A_2 \Big |_T)^\mu =&  \{v_1,v_2\}^\mu ~, \\
(A_1 \ast A_2 \Big |_{T^* \otimes T^*})_{\mu \nu} =&   ({\cal L}_{v_1} \hat \gamma_2   - {\cal L}_{v_2} \hat \gamma_1)_{\mu \nu}  + \nabla_{(\mu} ({\cal L}_{v_1} \rho_2   + 2  \ui_{v_2} \hat \gamma_1 )_{\nu)} ~, \\ 
(A_1 \ast A_2 \Big |_{T^*})_\mu =&    ({\cal L}_{v_1} \rho_2   +  2 \ui_{v_2}  \hat \gamma_1 )_\mu  ~,
\end{align}
\end{subequations}
where we have defined $(\hat \gamma_i)_{\mu \nu} \equiv (\gamma_i)_{\mu \nu} - \nabla_{(\mu} \rho_{i \nu )}$.
The symmetrization is defined with a normalization factor, \eg: $\gamma_{(\mu \nu)} \equiv \tfrac{1}{2} (\gamma_{\mu \nu}  + \gamma_{\nu \mu} )$.
 This operation $*$ satisfies the Leibniz identity!
In \eqref{ddjdkkkk} the anomalous term is given by
\begin{multline}
\Lambda(\epsilon_1,\epsilon_2; A_1, A_2) = i \intsloop  \epsilon_2 \partial \epsilon_1 \;   2 \bigl(
( \ui_{v_1} \gamma_2  + \ui_{v_2} \gamma_1)_\mu + \nabla_\mu v_1^\rho \rho_{2\rho}  - v_2^\rho \nabla_\mu \rho_{1\rho}  \bigr) \partial \Phi^\mu  \\
+ i \intsloop  \epsilon_2 \partial^2 \epsilon_1 \;   \bigl( \ui_{v_1} \rho_2 - \ui_{v_2} \rho_1 \bigr)~.
\end{multline}

 Let us show how this calculation gives rise to another example of a weak Dorfman-Courant algebra. 
  Let ${\cal E} = \Gamma(TM \oplus S^2 T^*M \oplus T^*M)$ be the space of vector fields plus symmetric tensors of second rank plus one forms
   on $M$.  Define ${\cal R} = \Gamma(T^*M)$. The symmetric bilinear form $\langle~,~\rangle : {\cal E} \otimes {\cal E} \rightarrow {\cal R}$
    is given by the expression
\begin{equation}
\langle v_1 + \gamma_1 + \rho_1, v_2 + \gamma_2 + \rho_2 \rangle =   ({\cal L}_{v_1} \rho_2 +  {\cal L}_{v_2} \rho_1  +  2 \ui_{v_2}  \hat \gamma_1     +  2 \ui_{v_1}  \hat \gamma_2 ) ~.
\end{equation}
Next, define a map $\ud_h : {\cal R} \rightarrow {\cal E}$, which sends a one-form to a symmetric tensor of second rank and 
 a one form, as follows:
\begin{equation}
\ud_h \alpha= \frac{1}{2} \nabla_{\mu}  \alpha_{\nu} (\ud x^\mu \otimes \ud x^\nu + \ud x^\nu \otimes \ud x^\mu) +  \alpha_\mu \ud x^\mu~,
\end{equation}
 where $\alpha= \alpha_\mu \ud x^\mu$ is a one form. 
 With these definitions, the symmetrization of the bracket (\ref{eq:symtensorbracket})  is given by
\begin{equation}  
  A_{1} * A_{2} + A_{2} * A_{1} =   \ud_h \langle   A_{1} ,   A_{2} \rangle ~.
\end{equation}
Moreover, one can check the property 
\beq
 \langle A, \ud_h \langle B, C\rangle \rangle = \langle A * B, C \rangle + \langle B, A* C \rangle~,
\eeq{dkdkd93939sks}
 where $A,B,C \in {\cal E}$. 
Thus, we conclude that with the present definitions, $({\cal E}, {\cal R}, \ud_h, \langle~,~\rangle, *)$ is a weak Courant-Dorfman 
  algebra.   This structure depends on the choice of a torsionless connection on $TM$. 

\section{Summary} 
\label{summary}

In this  work, we have investigated some of the algebraic properties of the algebra of Poisson brackets 
 between local functionals. We were considering the cotangent bundle of the (super) loop space 
  $T^*{\cal L}M$ as basis for our consideration. We have argued that there exist a natural 
    structure of weak Courant-Dorfman algebras. We have considered four different examples which 
     gave interesting specific cases of weak Courant-Dorfman algebras, realized on geometric objects of $M$.  
 One can generate infinity many more examples along those lines. For example, we can consider local 
  functionals of the form
\beq
J_\epsilon (A) = \intsloop \epsilon \biggl( v^\mu S_\mu + 
  \hspace{-2.5em}  \sum_{\hphantom{fixed} k_1+ \cdots + k_p = \text{fixed} }  \hspace{-2.0em}  A (\Phi)_{{\nu_1} ... {\nu_p}}  \uD^{k_1} \Phi^{\nu_1}  \ldots \uD^{k_p} \Phi^{\nu_p} \biggr ) ~,
  \eeq{sks003022}
which will give rise to Poisson subalgebras. In (\ref{sks003022}), the components of $A$ are not tensors in general. 
 Everything can be covariantized by  introducing a connection, as in  the fourth example in section \ref{newcurrents}. 
  For these functionals, one can repeat the analysis we have presented and get new
   weak Courant-Dorfman structures. 

There is also another interesting  aspect of the current work. The closed subalgebras under the Poisson bracket of local functionals can be interpreted as first class constrains and thus related to gauge symmetries of a theory. 
 This is exactly the case in our analysis when the anomalous term vanishes. 
  In \cite{Alekseev:2004np} some examples of theories corresponding to first class constraints of the form \eqref{bosoniccurrent} were presented.  It would be interesting to study the gauge symmetries which arise from the closed 
   algebras considered here.

 Let us comment on the wider context for our results.  The observations presented in this work are closely 
  related to  similar statements which have appeared previously in the literature. 
  Historically, the first reference goes back to 1980 when Gel'fand and Dorfman \cite{gelfand} developed 
   the framework for variational calculus with application to 
 integrable systems. From these considerations the Dorfman brackets initially appeared. 
The Dorfman bracket also appeared  in the context of  the chiral de Rham 
    complex (a sheaf of vertex algebras) in  \cite{bressler}. In a sense, the Alekseev-Strobl result is a classical version 
     of the observation in   \cite{bressler}. 
   We believe that these different observations are closely related.
 We think that the natural mathematical framework for a better understanding of our observations 
   would be the sheaf of (super)Poisson vertex algebras over $M$. 
  We hope to come back to this problem elsewhere. 
    
Another point we would like to stress is that the considerations presented in section \ref{Linfinity} easily can be 
 generalized for a  higher dimensional classical field theory. Indeed, a specific example has 
  been discussed previously in \cite{Bonelli:2005ti}, \cite{Guttenberg:2006zi}.  Thus, the appearance of a Leibniz algebra and 
   other related structures is a generic feature of field theory. Hopefully, it may help to understand 
    the quantization in a more algebraic way, in analogy with vertex algebras. 
   
\section*{Acknowledgement}
We thank Reimundo Heluani, Sebastian Guttenberg and  Dmitry Roytenberg for discussions.   
 We thank the program "Geometrical Aspects of String Theory" at  
  Nordita  where part of this work was carried out.  The research of M.Z. was supported by VR-grant 621-2008-4273.
   We thank the referee for pointing out mistakes in the previous version of the paper.

\appendix
\Section{Courant-Dorfman algebra}
\label{appendix}
In this appendix, we collect the standard properties of the Dorfman and Courant brackets for the reader's convenience. We also give the definitions of a Courant-Dorfman algebra, and of a weak Courant-Dorfman algebra.

On the section of tangent plus cotangent bundle $(TM\oplus T^*M)$ we can define 
 the canonical pairing $\Gamma(TM\oplus T^*M) \times \Gamma (TM\oplus T^*M)\rightarrow C^\infty(M)$
 \beq
  \langle v_1 + \beta_1, v_2 + \beta_2 \rangle =  (\ui_{v_1} \beta_2 + \ui_{v_2} \beta_1 )~,
 \eeq{jdjee0303}
 and the bilinear operation $\Gamma(TM\oplus T^*M) \times \Gamma (TM\oplus T^*M)\rightarrow  \Gamma (TM\oplus T^*M)$
\beq
   (v_1 + \beta_1) * (v_2 + \beta_2)= \{ v_1, v_2\} + {\cal L}_{v_1} \beta_2 - \ui_{v_2} d\beta_1~,
\eeq{definitionDorfman}
 which is called the Dorfman bracket. 
 In the above formulas $\{~,~\}$ is a Lie bracket of the vector fields, ${\cal L}_{v}$ is a Lie derivative, $\ui_v$ is a contraction and
  $\ud$ is the standard exterior derivative on the differential forms. 
  By a direct calculation one can check the Leibniz identity for the Dorfman bracket,
\begin{equation}\label{Leibniz}
   A * (B* C) = (A* B)*C + B* (A* C)~, 
 \end{equation}
$A, B, C \in \Gamma(TM\oplus T^*M)$, which makes $\Gamma(TM \oplus T^*M)$ into a Leibniz algebra.  Moreover, the following 
  additional properties are satisfied:
     \begin{subequations}
\begin{align}
\label{AA1}    A * (f B) &= f (A*B) + \langle A,  \ud f \rangle    B~,\\
\label{AA2}      A * B + B* A  &= \ud \langle A, B \rangle ~, \\
 \label{AA3}  \langle A, \ud \langle B, C\rangle \rangle &= \langle A * B, C \rangle + \langle B, A* C \rangle ~,\\
 \label{AA4}  \ud f  * A &=0~,\\
\label{AA5}   \langle \ud f, \ud g \rangle &=0~,
\end{align}
\end{subequations}
  where  $A, B, C \in \Gamma(TM\oplus T^*M)$ and $f, g \in C^\infty(M)$. 
  The Courant bracket  is defined as the antisymmetrization of the Dorfman bracket:
\beq
 [A, B]_C = \frac{1}{2} \left ( A*B - B*A\right )~.
\eeq{definitionCourant} 
As follows from (\ref{AA2}) that the Courant bracket is related to the Dorfman bracket as
\beq
 [A, B]_C = A*B -  \frac{1}{2} \ud \langle A, B\rangle~.
\eeq{CourantDorfman}
  If one understands the exterior derivative $\ud$ as a map $\ud : C^\infty(M) \rightarrow \Gamma(TM\oplus T^*M)$,
    then one can formalize the present structure through the notion of a Courant-Dorfman algebra. 
   
Inspired by the example of $TM\oplus T^*M$ and following \cite{roytenbergcour} , a {\it Courant-Dorfman algebra} 
 $({\cal E}, {\cal R}, \partial, \langle~,~\rangle, *)$
 consists of the following data:
\begin{enumerate}
\addtolength{\itemsep}{-0.25\baselineskip}
\renewcommand{\labelenumi}{(\alph{enumi})}
\setlength{\itemindent}{4em}
\item a commutative $\mathbb{K}$-algebra ${\cal R}$
 \item an ${\cal R}$-module ${\cal E}$
\item  a symmetric bilinear form $\langle~,~\rangle: {\cal E} \otimes {\cal E} \rightarrow {\cal R}$
 \item a derivation $\partial : {\cal R} \rightarrow {\cal E}$
 \item a Dorfman bracket $*: {\cal E} \otimes {\cal E} \rightarrow {\cal E}$
 \end{enumerate}
  which satisfy the following axioms:
\begin{enumerate}
\addtolength{\itemsep}{-0.25\baselineskip}
\renewcommand{\labelenumi}{(\arabic{enumi})}
\setlength{\itemindent}{4em}
 \item $A * (f B) = f (A * B) + \langle A, \partial f\rangle B $
 \item $\langle A, \partial \langle B, C\rangle \rangle = \langle A * B, C \rangle + \langle B, A* C \rangle $
 \item $ A * B + B* A = \partial \langle A, B \rangle  $
 \item $ A * (B* C) = (A* B)*C + B* (A* C) $
 \item $ (\partial f) * A = 0 $
 \item $ \langle \partial f, \partial g \rangle =0 $
 \end{enumerate}
   where $A,B,C \in {\cal E}$ and $f,g \in {\cal R}$.   
   
   If the symmetric bilinear form $\langle~,~\rangle$ is non-degenerate in an appropriate sense \cite{roytenbergcour},
     then  in the above definition the conditions (1), (5) and (6) are redundant. 
  
  In the present work, we need a weaker notion of the Courant-Dorfman algebra where ${\cal R}$ would not be 
   a commutative  algebra and ${\cal E}$ would not be an ${\cal R}$-module.  
 A   {\it weak Courant-Dorfman algebra}  $({\cal E}, {\cal R}, \partial, \langle~,~\rangle, *)$ is defined by 
   the following data:
\begin{enumerate}
\addtolength{\itemsep}{-0.25\baselineskip}
\renewcommand{\labelenumi}{(\alph{enumi})}
\setlength{\itemindent}{4em}
\item a vector space ${\cal R}$ 
\item a vector space ${\cal E}$
\item   a symmetric bilinear form $\langle~,~\rangle: {\cal E} \otimes {\cal E} \rightarrow {\cal R}$
\item  a map $\partial : {\cal R} \rightarrow {\cal E}$
\item  a Dorfman bracket $*: {\cal E} \otimes {\cal E} \rightarrow {\cal E}$
 \end{enumerate}
  which satisfy the following axioms:
\begin{enumerate}
\addtolength{\itemsep}{-0.25\baselineskip}
\renewcommand{\labelenumi}{\em{(\arabic{enumi})}}
\setlength{\itemindent}{4em}
\renewcommand{\labelenumi}{(\arabic{enumi})}
 \item   $  A * (B* C) = (A* B)*C + B* (A* C) $
 \item   $A * B + B* A = \partial \langle A, B \rangle  $
 \item   $ (\partial f) * A = 0 $
 \end{enumerate}
   where $A,B,C \in {\cal E}$ and $f,g \in {\cal R}$.   
    Comparing with the definition of a Courant-Dorfman algebra, the  properties related 
     to the algebraic structures of ${\cal R}$ and ${\cal E}$ has changed, and axiom 1, 2 and 6 are removed. Note that properties similar to axiom 2 and 6 of a Courant-Dorfman algebra follows from the axioms:
   \begin{equation}
   \partial \left ( \langle A, \partial \langle B, C\rangle \rangle - \langle A * B, C \rangle - \langle B, A* C \rangle\right )=0 ~, \\
   \end{equation}
   \begin{equation}
 \partial\left (  \langle \partial f, \partial g \rangle \right ) =0 ~.
   \end{equation}
  
  An example of a weak Courant-Dorfman algebra is given by 
   ${\cal E} = \Gamma(TM \oplus \wedge^p T^*M)$ and ${\cal R} = \Gamma(\wedge^{p-1}T^*M)$, where the symmetric bilinear form 
    $\langle~,~\rangle$ and the Dorfman bracket $*$ are  defined formally by the same formulas (\ref{jdjee0303}) and (\ref{definitionDorfman}) 
     but now $(v_1+\beta_1), (v_2 + \beta_2) \in \Gamma(TM \oplus \wedge^p T^*M)$.   The map $\partial$ is the exterior
       derivative acting on $(p-1)$-forms.
      All the properties can easily be checked 
      explicitly.


\begin{thebibliography}{6666}

\newcommand{\np}{{\em Nucl.\ Phys.\ }}
\newcommand{\pr}{{\em Phys.\ Rev.\ }}
\newcommand{\cmp}{{\em Commun.\ Math.\ Phys.\ }}
\newcommand{\pl}{{\em Phys.\ Lett.\ }}
%
\bibitem{AlAshhab:2003th}
  S.~Al-Ashhab,
  ``A Class of Strongly Homotopy Lie Algebras with Simplified sh-Lie
  Structures,''
  J. Pure Appl. Algebra 208 (2007), no. {\bf 2}, 647--653
  [arXiv:math/0308160].
%
\bibitem{Alekseev:2004np}
  A.~Alekseev and T.~Strobl,
  ``Current algebra and differential geometry,''
  JHEP {\bf 0503} (2005) 035
  [arXiv:hep-th/0410183].
%
\bibitem{Barnich:1997ij}
  G.~Barnich, R.~Fulp, T.~Lada and J.~Stasheff,
  ``The sh Lie structure of Poisson brackets in field theory,''
  Commun.\ Math.\ Phys.\  {\bf 191} (1998) 585
  [arXiv:hep-th/9702176].
%
\bibitem{Bonelli:2005ti}
  G.~Bonelli and M.~Zabzine,
  ``From current algebras for p-branes to topological M-theory,''
  JHEP {\bf 0509} (2005) 015
  [arXiv:hep-th/0507051].
%
\bibitem{Bredthauer:2006hf}
  A.~Bredthauer, U.~Lindstr\"om, J.~Persson and M.~Zabzine,
  ``Generalized K\"ahler geometry from supersymmetric sigma models,''
  Lett.\ Math.\ Phys.\  {\bf 77} (2006) 291
  [arXiv:hep-th/0603130].
%
\bibitem{bressler}
P.~Bressler,
 ``The first Pontryagin class,"
  Compos.\ Math.\ {\bf 143} (2007), no. 5
[arXiv:math.AT/0509563].
%
\bibitem{dickeybook} 
L.~A.~Dickey,
 ``Soliton equations and Hamiltonian systems,"
  Advanced Series in Mathematical Physics, vol. {\bf 12} Singapore: World Scientific, 1991.
%
\bibitem{dickeypaper}
L.~A.~Dickey,
 ``Poisson brackets with divergence terms in field theories: three examples,"
 Higher homotopy structures in topology and mathematical physics (Poughkeepsie, NY, 1996), 67--78, 
Contemp. Math., {\bf 227}, Amer. Math. Soc., Providence, RI, 1999. 
%
\bibitem{FigueroaO'Farrill:2005uz}
  J.~M.~Figueroa-O'Farrill and N.~Mohammedi,
  ``Gauging the Wess-Zumino term of a sigma model with boundary,''
  JHEP {\bf 0508} (2005) 086
  [arXiv:hep-th/0506049].
%
\bibitem{gelfand}
I.~M.~Gel'fand, I.~Ya.~Dorfman,
``Schouten bracket and Hamiltonian operators,"
 Funkt.nal. i Prilozhen. {\bf 14} (1980) 71-74;
 Funct.Anal. Appl. {\bf 14} (1981), 223-226. 
%
\bibitem{Guttenberg:2006zi}
  S.~Guttenberg,
  ``Brackets, sigma models and integrability of generalized complex
  structures,''
  JHEP {\bf 0706} (2007) 004
  [arXiv:hep-th/0609015].
%
\bibitem{Derived}
Y.~Kosmann-Schwarzbach,
``Derived Brackets,"
Lett. in Math. Phys., vol.{\bf  69}, 61-87
 [arXiv:math/0312524].
%
\bibitem{markl}
M.~Markl and S.~Shnider,  
 ``Differential operator endomorphisms of an Euler-Lagrange complex,"
 Tel Aviv Topology Conference: Rothenberg Festschrift (1998),  177--201, Contemp. Math., {\bf 231}, 
 Amer. Math. Soc., Providence, RI, 1999
 [arXiv:math.DG/9808105].
%
\bibitem{roytenbergweinst}
D.~Roytenberg, A.~Weinstein,
  ``Courant Algebroids and Strongly Homotopy Lie Algebras,"
   Lett.\ Math.\ Phys.\  {\bf 46} (1998) 81
  [math.QA/9802118].
%
\bibitem{roytenbergcour}
D.~Roytenberg,
 ``Courant-Dorfman algebras and their cohomology."
 arXiv:0902.4862.
%
\bibitem{Zabzine:2006uz}
  M.~Zabzine,
  ``Lectures on generalized complex geometry and supersymmetry,''
 	Archivum mathematicum (supplement) {\bf 42}:119-146,2006
  [arXiv:hep-th/0605148].

  \end{thebibliography}
\end{document}